\documentclass[onecolumn]{aastex6}
\usepackage{amsmath}
\usepackage{color,txfonts}
\usepackage{hyperref}
\usepackage{epstopdf}
\usepackage{graphicx}
\usepackage[FIGTOPCAP]{subfigure}
\usepackage{amssymb}

\begin{document}

\title{GW170817: The energy extraction process of the off-axis relativistic outflow and the constraint on the equation of state of neutron stars}
\author{Yuan-Zhu Wang$^{1,3}$, Dong-Sheng Shao$^{1,2}$, Jin-Liang Jiang$^{1,2}$, Shao-Peng Tang$^{1,2}$, Xiao-Xiao Ren$^{1,2}$, Fu-Wen Zhang$^{4}$, Zhi-Ping Jin$^{1,2}$,
	Yi-Zhong Fan$^{1,2}$, and Da-Ming Wei$^{1,2}$}
\affil{
	$^1$ {Key Laboratory of dark Matter and Space Astronomy, Purple Mountain Observatory, Chinese Academy of Science, Nanjing, 210008, China.}\\
	$^2$ {School of Astronomy and Space Science, University of Science and Technology of China, Hefei, Anhui 230026, China.}\\
	$^3$ {University of Chinese Academy of Sciences, Yuquan Road 19, Beijing, 100049, China}\\
	$^4$ {College of Science, Guilin University of Technology, Guilin 541004, China.}\\
}
\email{wangyz@pmo.ac.cn(YZW), yzfan@pmo.ac.cn (YZF) and dmwei@pmo.ac.cn (DMW)}
	
\begin{abstract}
As revealed recently by the modeling of the multi-wavelength data of the emission following GW170817/GRB 170817A, there was an off-axis energetic relativistic outflow component launched by this historic double neutron star merger event. In this work we use the results of these modeling to examine the energy extraction process of the central engine. We show that the magnetic process (i.e., the Blandford-Znajek mechanism) is favored, while the neutrino process usually requires a too massive accretion disk if the duration of the central engine activity is comparable to the observed $T_{90}$ of GRB 170817A, unless the timescale of the central engine activity is less than $\sim$ 0.2s. We propose that the GRB observations are helpful to constrain the combined tidal parameter $\tilde{\Lambda}$, and by adopting the accretion disk mass distribution estimated in BZ mechanism, the $90\%$ credible interval of $\tilde{\Lambda}$ for the progenitor of GW170817 is inferred as $309-954$.
\end{abstract}

\section {Introduction}
After the successful detection of the first binary neutron star coalescence event (GW170817; \citep{2017PhRvL.119p1101A}), a large number of follow-up observations have been carried out  \citep[e.g.,][]{2017ApJ...848L..21A,2017Sci...358.1556C,2017NatAs...1..791C,2017ApJ...848L..17C,2017ApJ...848L..14G,2017ApJ...848L..20M,2017Natur.551...67P,2017ApJ...848L..15S,2017Natur.551...71T}, leading to a broad discussion on the physical picture underlying the multi-wavelength signals. In the field of short gamma-ray burst (sGRB), the energy flux of GRB 170817A, the observed inclination angle of the binary orbital axis with respect to our line of sight (although with a large uncertainty) as well as the behaviors of the X-ray and radio afterglow do help to constrain the GRB models \citep{2017ApJ...848L..13A,2018ApJ...857..128J,2018PhRvL.120x1103L,2018ApJ...860L...2F,2018MNRAS.476.1785B,2018ApJ...853L..10Y,2019arXiv190101521D}; however, the central engine and how it powered the multi-wavelength emission is still in debate \citep[][see also \citet{2013ApJ...779L..25F} for a highly relevant investigation though on GRB 130603B, the first short event with a reliably identified macronova/kilonova signal that should have a neutron star merger origin too]{2017Sci...358.1559K,2017ApJ...848L..34M,2018ApJ...861..114Y,2018ApJ...858...74M,2018ApJ...860...72M,2018Natur.561..355M,2018ApJ...856L..33G,2018ApJ...856...90L,2018ApJ...859L..23P,2018ApJ...859L...3L,2018NatCo...9..447Z}.

Long before the era of gravitational wave astronomy, three competing models for sGRBs' central engine energy extraction have been proposed and widely discussed, including the Blandford-Znajek (BZ hereafter) mechanism \citep{1977MNRAS.179..433B} or neutrino annihilation \citep{1989Natur.340..126E} of the BH-accretion disk system, and the model involving a millisecond magnetar central engine \citep{Duncan1992,2006ChJAA...6..513G}. In the BH-accretion disk scenarios, one widely known fact is that the Blandford-Znajek mechanism can launch the relativistic outflow much more efficiently than the neutrino process when the accretion rate is lower than $\sim 0.01M_\odot~{\rm s}^{-1}$ \citep{2005ApJ...635L.129F}. If the progenitor stars of a sGRB is a neutron star binary, benefited from the narrowly distributed total gravitational mass ($M_{\rm tot}$) of the Galactic binary neutron star systems and the reasonably evaluated dimensionless spin parameter of the nascent black hole, the mass of the accretion disk ($M_{\rm disk}$) that launched the sGRB outflow can be straightforwardly estimated with the observation data, as initially proposed by \citet{2011ApJ...739...47F}. These authors then collected 10 sGRBs with the relatively reasonably-constrained geometry-corrected energy outputs. Within the neutrino scenario, the required $M_{\rm disk}$ are found to be too massive to be realistic in half of the sample and the magnetic process is needed \citep{2011ApJ...739...47F}.  \citet{2015ApJS..218...12L} compared the predicted luminosities of the two mechanisms with both long and short burst samples, and their result showed that with the assumed combinations of spin and disk mass, the neutrino annihilation may be unable to produce enough luminosity for a good fraction of the events. The magnetic process was also favored in accounting for the tentative GBM Transient 150914 \citep{2016ApJ...827L..16L} that was claimed to be potentially associated with GW150914 \citep{2016ApJ...826L...6C}.

However, for GRB 170817A, the first short burst with a detected gravitational wave signal, the approach of \citet{2011ApJ...739...47F} can not be directly applied. As an extremely under-luminous short event, the physical origin of GRB 170817A itself is still unclear. Fortunately, the gravitational wave data \citep{2017PhRvL.119p1101A} and in particular the
very long-baseline interferometry (VLBI) data \citep{2018Natur.561..355M} unambiguously suggest that GW170817/GRB 170817A are off-axis events. Therefore, it is most likely that the relativistic outflow of GW170817/GRB 170817A is structured (for example, a narrowly-collimated ultra-relativistic jet with a wide mildly-relativistic outflow/cocoon, as suggested in \citet{2007ApJ...656L..57J,2018ApJ...857..128J}), for which the $M_{\rm disk}$ can not be reliably estimated with the data of GRB 170817A alone. The modeling of the multi-wavelength afterglow data well constrain the total energy of the off-axis relativistic jet \citep{2018Natur.561..355M,2018arXiv180806617V,2019ApJ...870L..15L}. If this energy is extracted from a rapidly rotating black hole (BH) formed in the merger, then $M_{\rm disk}$ can be inferred and could help to clarify the appropriate energy extraction process. \citet{2018ApJ...869..130R} found a connection between the combined tidal deformability parameter $\tilde{\Lambda}$ and the disk mass, and with this relation, the inferred disk mass can be further used to constrain the equation of state (EoS) of neutron stars, as revealed in \citet{2018ApJ...852L..29R}. This work is mainly motivated by such prospects.

The structure of our work is as follows. In Section 2, by assuming the central remnant of GW170817 is a black hole (BH) with a hyper-accreting disk, the prospects of BZ mechanism and neutrino annihilation process for launching relativistic outflow of GW170817/GRB 170817A are investigated based on Bayesian inference, and the $M_{\rm disk}$ distribution for BZ mechanism is derived. The constraint on the EoS from GW/GRB joint observations is examined in Section 3. We have some discussions in Section 4.

\section{The disk masses estimation: the neutrino annihilation model and BZ mechanism}
The roles of the neutrino annihilation and BZ mechanism in launching the relativistic outflow from a hyper-accreting black hole have been widely investigated \citep[][and the references therein]{2015ApJS..218...12L}.  For GW170817, the total gravitational mass of the BNS $\sim 2.75 \rm{M_{\odot}}$ is larger than 1.2 times of the TOV mass of many EoSs, so it is very likely that a hypermassive neutron star (HMNS) was formed (a prompt BH could be formed for some soft EoSs). The HMNS was short lived since the viscosity effect, GW radiation and neutrino cooling usually brake the differential rotation quickly. The lifetime of HMNS can range from several milliseconds to greater than  tens of milliseconds \citep{2017PhRvD..96l3012S,2018ApJ...869..130R}. Since the delay of GRB 170817A with respect to the merger time is $\sim 1.75 \rm{s}$, we assume that the nascent HMNS had collapsed earlier than that.

The energy output of both BZ and neutrino (annihilation) mechanisms are related to the accretion disk mass ($M_{\rm disk}$). Below we briefly introduce the calculation formulae, compare the two models with Bayesian approaches and generate the possibility distributions of $M_{\rm disk}$ through Monte Carlo simulations.

\subsection{neutrino annihilation model}
The energy output of a hyper-accreting stellar mass BH via the neutrino-antineutrino annihilation process has been widely discussed in the literature  \citep[e.g.,][]{1999ApJ...524..262M,1999ApJ...518..356P,2007ApJ...661.1025L,2011MNRAS.410.2302Z}. With the empirical relation between the annihilation luminosity and the parameters of central engine found by \citet{2011MNRAS.410.2302Z}, \citet{2011ApJ...739...47F} proposed that in the double neutron star merger-powered on-axis sGRB scenario the disk mass can be estimated as
\begin{equation}\label{eq:md(na)}
\begin{aligned}
m_{\rm disk}\approx&0.735 M_\odot \left[\frac{E_{\rm GRB,51}}{{\mathcal F_{\rm grb}}}\right]^{4/9}\\
&\times\left(x_{\rm ms}\over 1.45\right)^{2.1}\left(\frac{M_{\rm BH}}{2.7M_\odot}\right)^{2/3}\left(\frac{T_{\rm act}}{\rm 1s}\right)^{5/9},
\end{aligned}
\end{equation}
where $E_{\rm GRB,51}$ is total geometry-corrected outflow energy of the on-axis GRB scaled by $10^{51} \rm{erg}$, $x_{\rm ms}=r_{\rm ms}/r_{\rm g}$ ($r_{\rm ms}$ is the radius of the last stable orbit, $r_{\rm g}$ is the Schwarzchild radius), $T_{\rm act}$ is the duration of the central engine activity, and $M_{\rm BH}$ is the mass of black hole. $\mathcal F_{\rm grb}$ is the fraction of the total neutrino annihilation energy that eventually powered the GRB outflow. We note that the above equation actually estimates the infalling mass of the accretion disk, which accounts for the majority of the total disk mass $M_{\rm disk}$. Taking into account the mass carried away by the wind as $m_{\rm wind}$, we have $M_{\rm disk} = m_{\rm disk} + m_{\rm wind}$.

\subsection{Blandford-Znajek mechanism}
In the presence of an ordered strong magnetic field in the accretion disk, the BZ mechanism \citep{1977MNRAS.179..433B} would possibly be the dominant way of extracting power from the rapidly rotating black hole. The total luminosity of the relativistic outflow driven by this mechanism is \citep{2000PhR...325...83L}
\begin{equation}
L_{\rm BZ}\approx2.5\times10^{49} {\rm erg\ s^{-1}} (a/0.5)^2(M_{\rm BH}/2.7M_\odot)^2B_{\rm H,15}^2,
\end{equation}
where $a$ is the dimensionless spin parameter of the BH, $B_{\rm H}\sim1.1\times10^{15} {\rm G} (\dot m/0.01 M_\odot~{\rm s}^{-1})^{1/2}R_{H,6}^{-1}$ is the strength of the magnetic field in the horizon, $\dot{m}$ is the accretion rate and $R_{\rm H}=(1+\sqrt{1-a^2})r_{\rm g}/2=(1+\sqrt{1-a^2})GM_{\rm BH}/c^{2}$, then we have
\begin{equation}
L_{\rm BZ}\approx 7.56\times10^{50} {\rm erg\ s^{-1}} \left(\frac{a}{1+\sqrt{1-a^2}}\right)^2(\dot m/0.01 M_\odot~{\rm s}^{-1}).
\end{equation}
Multiplying $T_{\rm act}$ in both sides and considering that $E_{\rm GRB}\approx {\mathcal F_{\rm grb}}L_{\rm BZ}T_{\rm act}$ and $m_{\rm disk}=\dot m T_{\rm act}$, we have
\begin{equation}\label{eq:md(bz)}
m_{\rm disk}\approx 0.0132M_\odot {1\over {\mathcal F_{\rm grb}}}\frac{E_{\rm GRB}}{10^{51}~{\rm erg}}\left(\frac{1+\sqrt{1-a^2}}{a}\right)^2.
\end{equation}
Interestingly, the disk mass is independent of the BH mass and the duration of the central engine.

\subsection{Bayesian Model Selection}
Above we have outlined the relationships between $M_{\rm disk}$ and the total energy of relativistic outflow as well as other relevant parameters for the given mechanisms. The two mechanisms usually predict different disk masses for the same parameter set. It is interesting to investigate which mechanism plays a leading role in the case of GRB 170817A. Previous studies have shown that the BZ mechanism is more efficient than the neutrino process, and the later one is incapable of powering some bursts with large total energy \citep{2011ApJ...739...47F,2015ApJS..218...12L}.

By adopting four different EoSs, \citet{2018ApJ...869..130R} found a correlation between disk mass and the combined tidal parameter $\tilde{\Lambda}$, i.e.,
\begin{equation}\label{eq:md(lambda)}
\frac{M_{\rm disk}}{M_{\odot}} = \max\left \{ 10^{-3},\bar{\alpha} + \beta \tanh ( \frac{\tilde{\Lambda} - \gamma}{\delta} )\right \}
\end{equation}
where $\bar{\alpha}=0.084$, $\beta = 0.127$, $\gamma = 567.1$, and $\delta= 405.14$. Since $\tilde{\Lambda}$ of the BNS, in principle, can be derived with the gravitational wave data, which can then be adopted to constrain $M_{\rm disk}$. Indeed, $\tilde{\Lambda}$ for GW170817 has already been inferred from the LIGO/Virgo observation data, which can be used to constrain the mass of the accretion disk through Eq.(\ref{eq:md(lambda)}) and then help us to distinguish between the energy extraction models. We collect the posterior samples of the BNS parameters from \citet{2019PhRvX...9a1001A} (here we only consider the low-spin case), and calculate the posterior distribution of the combined tidal parameter $\tilde{\Lambda}(\Lambda_1,\Lambda_2,m_1,m_2)$. The resulting histogram of $\tilde{\Lambda}$ is then smoothed with kernal density estimation (KDE), and we re-sample $10^{4}$ $\tilde{\Lambda}$ to calculate the posterior distribution of $M_{\rm disk}$.

\begin{figure}[ht!]
	\figurenum{1}\label{fig:lambdamass}
	\centering
	\includegraphics[angle=0,scale=1.0]{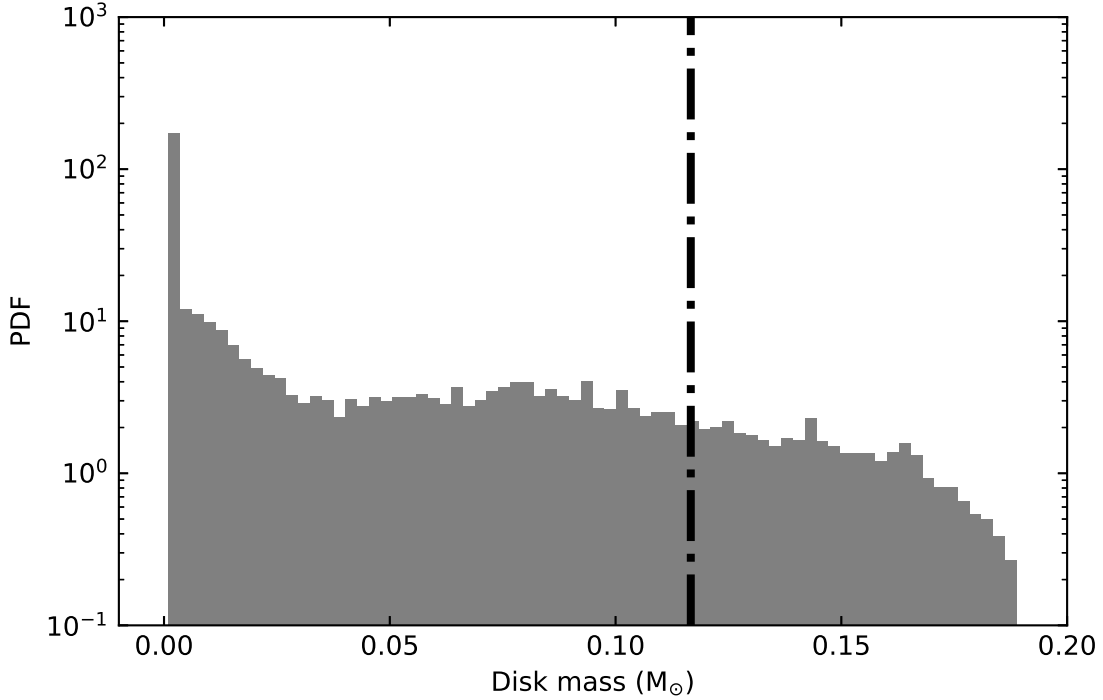}
	\caption{The normalized histogram for the simulated $M_{\rm disk}$ derived from the posterior distributions of $\tilde{\Lambda}$ constrained by Ligo/Virgo observation. The thick black vertical lines marks its 90th percentiles. Note that the y-axis is in log scale.}
	\hfill
\end{figure}

In Fig.\ref{fig:lambdamass}, we present the histogram of $M_{\rm disk}$ based on the constraint of advanced LIGO/Virgo on $\tilde{\Lambda}$. A large portion of the samples ($\sim 40\%$) concentrates in the first bin (i.e., $M_{\rm disk} \sim 0.001 M_\odot$). This is understandable as the peak of $\tilde{\Lambda}$ posterior distributions inferred from the LIGO/Virgo data is below 250 \citep{2019PhRvX...9a1001A}. With Eq.(\ref{eq:md(lambda)}), it is straightforward to show $M_{\rm disk} \lesssim 0.001 M_\odot$ for $\tilde{\Lambda}\leq 250$. Since the probability density function (PDF) is inversely proportional to $M_{\rm disk}$ around its minimum value, we can only place a $90\%$ upper limit on the $M_{\rm disk}$, which is $0.116 M_\odot$. We then use this information (i.e., the disk mass of the central remnant of GRB 170817A did not exceed $0.116 M_\odot$) as the criterion to compare the two models in the framework of Bayesian inference as follows:

First, we approximate the likelihood $\cal{L}$ for a giving parameter set $\theta$ as:
\begin{equation}
{\cal L}\left ( D \mid \theta \right ) = P\left [ M_{\rm disk}\left ( \theta \right ) < 0.116 M_\odot \right ]
\end{equation}
By assuming $\log(M_{\rm disk})$ to have a Gaussian distribution with mean $M_{\rm disk}(\theta)$, the likelihood can be expressed as:
\begin{equation}
\label{eq:pmd}
P\left [ M_{\rm disk}\left ( \theta \right ) < 0.116 M_\odot \right ] = 1-\frac{1}{2}\left [ 1-{\rm erf} \left ( \frac{\log\left ( 0.116 \right ) - M_{\rm disk}\left ( \theta \right )}{\sqrt{2}\sigma} \right )\right ]
\end{equation}
where the $\sigma$ represents the accuracy of Eq.(\ref{eq:md(na)}) or Eq.(\ref{eq:md(bz)}) on predicting the disk mass from the given parameter sets. This treatment is inspired by \citet{2018arXiv181012917R}, though we use the similar formulae for different purposes. Since it is difficult to investigate in detail about the accuracies of the two equations, we take different values of $\sigma$ in our calculation, and study how they affect our result.

Second, we assign priors for the uncertain parameters in Eq.(\ref{eq:md(na)}) and Eq.(\ref{eq:md(bz)}). For GRB 170817A, the estimated outflow energy in the literature \citep{2018Natur.561..355M,2018arXiv180806617V,2019ApJ...870L..15L} is based on the very late afterglow observation, and does not account for the prompt emission (the energy fluence of GRB 170817A measured by Fermi-GBM was recorded off-axis, which is likely well below the intrinsic energy of the prompt emission). We denote the outflow energy estimated via afterglow modeling and the intrinsic prompt gamma-ray energy as $E_{\rm k}$ and $E_\gamma$, then the $E_{\rm GRB}$ in Eq.(\ref{eq:md(na)}) and Eq.(\ref{eq:md(bz)}) can be expressed as $E_{\rm GRB}= E_{\rm k} + E_\gamma$. We adopt the parameters obtained from the afterglow modeling in \citet{2019ApJ...870L..15L} based on the latest observations to derive our prior for $E_{\rm k}$. Two configurations of the jet's structure are applied in their afterglow fitting: the gaussian configuration and the jet+cocoon configuration; we include both configurations in our analysis. Due to these parameters are reported in a way of $median^{+ 84th \ percentile}_{-16th \ percentile}$ instead of showing the detail of the distributions, we approximate their distribution through the method introduced by \citet{2013ApJ...778...66K} and calculate the $E_{\rm k}$ distribution. Being viewed off-axis, the total $E_\gamma$ of GRB 170817A is unkown; however, it should be reasonable to assume that the isotropic equivalent energy ($E_{\rm iso}$) of GRB 170817A is similar to the whole sGRB population if it was observed on-axis. We collect the information of sGRBs with redshift measured \citep{2012ApJ...750...88Z,2015ApJ...815..102F,2017ApJ...850..161T,2016ApJ...829....7L,2016ApJS..223...28N} to calculate their $E_{\rm iso}$, and their distribution is treated as the prior of $E_{\rm iso}$ for GRB 170817A. The prior of $E_\gamma$ is then calculated by correcting the $E_{\rm iso}$ with the half-openning angle derived in \citet{2019ApJ...870L..15L}.

It has been suggested that the blue component of the Kilonova AT2017gfo originated from the disk wind of the central remnant, so the modeling of kilonova light curve could reveal the information about $m_{\rm wind}$. We take the fitting result from \citet{2018arXiv181204803C} (the posterior for $m_{\rm ej2}$ in their paper) to construct the prior for $m_{\rm wind}$ using the same approach for the construction of prior for $E_{\rm k}$.

The spin of the final BH is found to be narrowly distributed \citep{2009PhRvD..80f4037K}. For prompt BH formation it is $\sim 0.78$, weakly depending on the total mass and the mass ratio of the BNS. If a HMNS initially formed, the spin of the final BH is slightly smaller, ranging from 0.59 to 0.76 in \citet{2017PhRvD..95b4029D} and \citet{2017PhRvD..96l3012S}. Considering there is no reliable measurement on the spin, we take a uniform distribution for $a$ ranging from 0.6 to 0.8.

The constraint on the bulk Lorentz factor of the outflow by \citet{2018Natur.561..355M} is $\Gamma_0 > 10$. \citet{2005A&A...436..273A} showed that the matters with Lorentz factors larger than 10 occupy $\sim 0.35-0.6$ of the total annihilation energy. Therefore, the distribution of $\mathcal F_{\rm grb}$ is taken as a uniform distribution within this range. We assume that the $\mathcal F_{\rm grb}$ for BZ mechanism also follows this distribution.

In the neutrino annihilation model, the mass of the central BH is needed. One information that can be included is the total mass of the BNS measured by Ligo/Virgo observation, which is $\sim 2.73 \rm{M_{\odot}}$ \citep{2019PhRvX...9a1001A}. Due to the energy conservation, the mass of the final black hole can be approximately estimated by \citep{2009PhRvD..80f4037K}:
\begin{equation}\label{eq:mg}
M_{\rm BH} = M_{\rm tot} - M_{\rm disk} -M_{\rm ej} - \Delta E\\
\end{equation}
where $M_{\rm ej}$ is mass of dynamical ejecta, $M_{\rm tot}$ represent the total gravitational mass of the BNS, and $\Delta E$ is the energy lose by gravitational wave, the change of binding energy during the formation and collapse of HMNS (if the BH is not promptly formed) and other dissipating processes. Generally speaking, the amount of ejected mass could not be larger than several $0.01 \rm{M_{\odot}}$ \citep{2017PhRvD..96l3012S, 2018ApJ...869..130R}, except for some very hard EOSs like MS1b \citep{2017PhRvD..95b4029D} that might have already been excluded by LIGO observation \citep{2017PhRvL.119p1101A}. We also assume that $\Delta E$ is small comparing to the total gravitational mass, thus we take $M_{\rm BH} \approx M_{\rm tot} - M_{\rm disk}$ for simplicity\footnote{We have checked that even for a large $\Delta E = 0.2 M_\odot$ will only lead a minor change on the final $M_{\rm disk}$ based on its relatively weak dependency on $M_{\rm BH}$, and all of the following conclusions still hold.}.

There is no direct measurement of the duration of the central engine activity for GRBs, and it is generally assume that this time scale is relevant to the duration of the prompt emission, i.e., $T_{\rm act} \lesssim T_{90}$. For GRB 170817A the off-axis observation makes the evaluation of $T_{\rm act}$ more challenging. In the following calculations we first assume that it is close to the \emph{Fermi} measured $T_{90}$ of 2 second, and further investigate how the result changes with different time scales.

Third, having the likelihood being defined for the parameters (i.e., $E_{\rm k}$, $E_{\gamma}$, $\mathcal F_{\rm grb}$, $m_{\rm wind}$, $a$) being assigned to their distributions, we compute the Bayesian evidence ($\mathcal{Z}$) for the two models:
\begin{equation}
\mathcal{Z} = \int \ d\theta\mathcal{L}\left ( D\mid\theta \right )\pi\left ( \theta \right )
\end{equation}
where $\pi(\theta)$ is the prior. The Bayes factor and odds ratio is then derived by:
\begin{equation}
\mathcal{O_{NA}^{BZ}} = \frac{\mathcal Z_{\rm BZ}}{\mathcal Z_{\rm NA}}\frac{\pi_{\rm BZ}}{\pi_{\rm NA}}
\end{equation}
We set the prior odds ratio to unity, so the Bayes factor $\mathcal{B}=\frac{\mathcal Z_{\rm BZ}}{\mathcal Z_{\rm NA}}$ is equals to odds ratio $\mathcal{O_{NA}^{BZ}}$, and is used to compare the two models. The calculation is performed using the nested sampling package MULTINEST \citep{2009CQGra..26u5003F}.

In Tab.1 we present the evidence $\ln(\mathcal{Z})$ and Bayes factors $\mathcal{B}$ between the two models calculated with different $\sigma$, by assuming $T_{\rm act}=2 \rm{s}$. Three different values of $\sigma$ (0.5, 0.3 and 0.1) are adopted in our calculation. For every $\sigma$, the BZ mechanism yields larger $\mathcal{Z}$ than neutrino annihilation, thus the bayes factor $\frac{\mathcal Z_{\rm BZ}}{\mathcal Z_{\rm NA}}$ is always larger than unity in all the comparisons. Accroding to \citet{K1995}, a Bayes factor lager than (3, 50, 150) gives (positive, strong, very strong) support for preferring the BZ model than neutrino annihilation model. We can know from Tab.1 that even for a very large $\sigma$ of 0.5 (which means that regarding the $68\%$ credible interval covers one order of magnitude), the data prefers the BZ model 9.0 and 6.7 times over the neutrino annihilation model for Gaussian and jet+cocoon structure respectively. The $\ln(\mathcal{Z})$ slightly decreases as the $\sigma$ increases for BZ mechanism, while it increases in the case of neutrino annihilation model. This may imply that for the majority of parameter space, the BZ mechanism predicts a disk mass below the limitation from LIGO/Virgo observation, while the neutrino annihilation model predicts a disk mass exceeding this limit.

\begin{table}
	\label{tab:results}
	\begin{center}
		\title{}Table 1. The evidences for the two models and the Bayes factors between them.\\
		\begin{tabular}{lllll} \hline \hline
			Sturcture	&  $\sigma$	& $ln\mathcal{Z_{\rm BZ}}$	&  $ln\mathcal{Z_{\rm NA}}$	& $\mathcal{B}$ \\  \hline \hline
		             	&    0.5    & -1.0                   	&  -3.2				        & 9.0       	\\
			Gaussian    &    0.3    & -0.9                   	&  -4.8			            & 49.4       	\\
		            	&    0.1    & -0.8                   	&  -9.5			            & 6002.9       	\\
			\hline
		            	&    0.5    & -0.9                   	&  -2.8				        & 6.7       	\\
			Jet+cocoon	&    0.3    & -0.7                   	&  -3.5				        & 16.4      	\\
		            	&    0.1    & -0.6                  	&  -5.1				        & 90.0       	\\
			\hline
		\end{tabular}
	\end{center}
	\textbf{Notes.}
	a. NA stands for neutrino annihilation, and BZ stands for Blandford-Znajek mechanism. \\
	b. The errors of evidence returned by MULTINEST are smaller than the last digit of reported values. \\
\end{table}

\begin{figure}[ht!]
	\figurenum{2}\label{fig:B_changes}
	\centering
	\includegraphics[angle=0,scale=1.0]{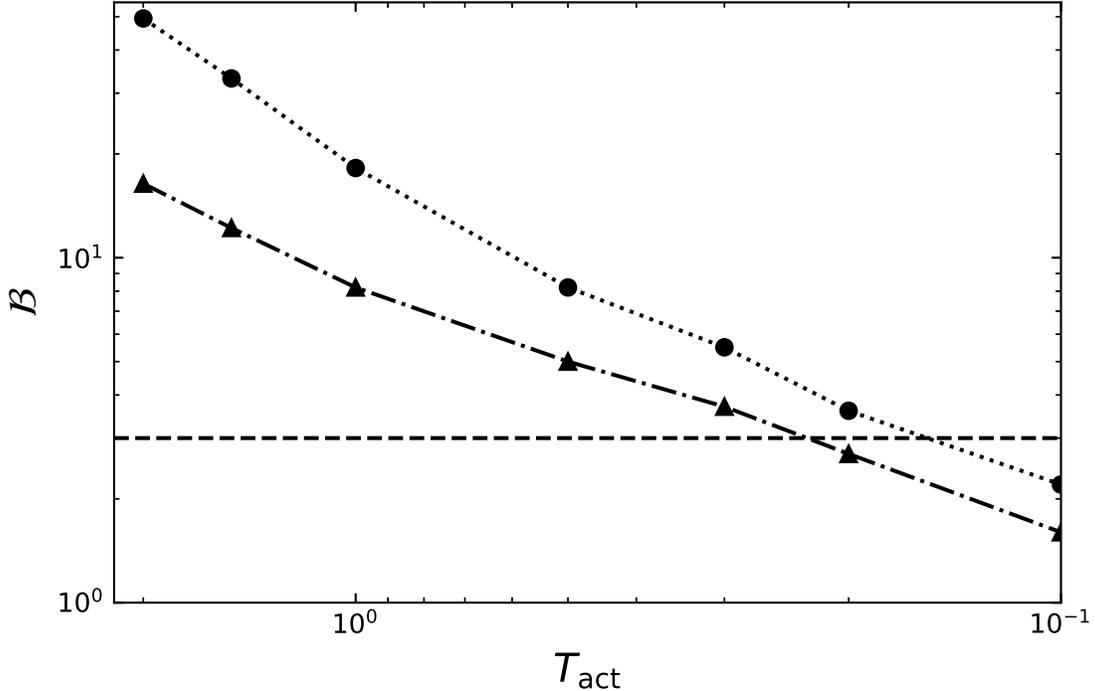}
	\caption{The Bayes factor as a functon of $T_{\rm act}$ for $\sigma = 0.3$. Results from the Gaussian structure are marked with solid circle, and results from the jet+cocoon structure are marked with solid triangle. the horizontal dashed line indecates $\mathcal{B} = 3$, below which the two model become indistinguishable.}
	\hfill
\end{figure}

Note that for neutrino annihilation model, if we just account for the dispersion of the prediction by the empirical function on the simulation results, one can find from \citet{2011MNRAS.410.2302Z} that within our region of interest, most simulated points have residuals smaller than 0.3 in log-space; the relevant equations for BZ mechanism originate from the analytical calculation by \citet{2000PhR...325...83L} and do not have this dispersion. We assume a $\sigma$ of 0.3 is conservative enough for Eq.(\ref{eq:md(na)}) and Eq.(\ref{eq:md(bz)}). In the above analysis, we assume that the time scale of the central engine activity is comparable to the \emph{Fermi} measured $T_{90}$ of the prompt emission. In Fig.\ref{fig:B_changes}, we show the Bayes factor as a function of $T_{\rm act}$ by taking $\sigma = 0.3$. The $\mathcal{B}$ decreases as $T_{\rm act}$ increases, and this is easy to understand from Eq.(\ref{eq:md(na)}) that the disk mass is proportional to $T_{\rm act}^{5/9}$. Interestingly, for both structure configuratons, the neutrino annihilation model become comparable ($\mathcal{B}<3$) with the BZ mechanism for $T_{\rm act}$ less than $\sim 0.2 {\rm s}$, which is about one order of magnitude lower than the $T_{90}$.

The only information we use to compute the likelihood is how likely the predicted disk mass is below the limitation from LIGO/Virgo observation, and due to the degeneracy of parameters, the parameter space are not well constrained for both models. Even so, the results from MULTINEST sampling show that for neutrino annihilation model the allowed parameter space under the limitation is much smaller than the one for BZ mechanism.
Therefore, we conclude that the limitation of $\tilde{\Lambda}$ from the observation of GW 170817A favors the BZ mechanism rather than the neutrino annihilation model (although it can not be convincingly ruled out), unless the timescale of the central engine activity is less than $\sim 0.2 {\rm s}$.

\subsection{The Disk Mass Distribution}

In the previous Section, we have used the constraint on $\tilde{\Lambda}$ set by the gravitational wave data to test the energy extraction models and found out that the BZ mechanism is favored. Below we assume that the BZ mechanism was the underlying process of launching the relativistic outflow following GW170817, and derive the disk mass distribution for the central remnant of GW170817/GRB 170817A.

We perform the Monte Carlo simulation to derive the probability distribution of $M_{\rm disk}$. We generate $10^5$ parameter sets from their prior distributions as described in the section above, and then calculate $M_{\rm disk}$ with Eq.(\ref{eq:md(bz)}). As mentioned before, the distributions of $E_{\rm \gamma}$, $E_{\rm k}$ and $m_{\rm wind}$ are determined from previous observations, whereas the distributions of $\mathcal{F}_{\rm grb}$ and $a$ are unclear. To cover the possible situations, considering the monotonic relation between $M_{\rm disk}$ and the two undetermined parameters, we perform the simulation in three different group: i. the $\mathcal{F}_{\rm grb}$ and $a$ are fixed to 0.6 and 0.6 respectively (noted as high mass group); ii. the distributions of $\mathcal{F}_{\rm grb}$ and $a$ are taken as same as the uniform prior distributions that are discussed in the previous section (noted as median mass group); iii. the $\mathcal{F}_{\rm grb}$ and $a$ are fixed to 0.35 and 0.8 respectively (noted as low mass group).

Our analysis is based on the assumption that the central engine of GRB 170817A is a BH$+$disk system. In general, the maximum gravitational mass of a fast spinning neutron star is believed to be $\sim 1.2$ times of $M_{\rm TOV}$, and since the maximal mass of a non-rotating neutron star is $\geq 2.01 M_{\odot}$, it is reasonable to speculate that if the gravitational mass of the remnant after merger is less than 2.42 $M_{\odot}$, it will not collapse into a Black Hole. Based on the conservative of baryon number, the change of gravitational mass between the total mass of BNS (which is measured by gravitational wave observation to be $2.73 M_{\odot}$) and the launch of the sub-relativistic outflow with the mass of $\sim 0.05 M_{\odot}$, thus if the mass of the disk exceeds $\sim 0.2 M_{\odot}$, the central remnant will not be a black hole, inconsistent with our initial assumption. To solve this problem we introduce an extra prior that $P(M_{\rm disk})=0$ for $M_{\rm disk} \geq 0.2 M_{\odot}$.

\begin{figure}[ht!]
	\figurenum{3}\label{fig:diskmass}
	\centering
	\includegraphics[angle=0,scale=0.5]{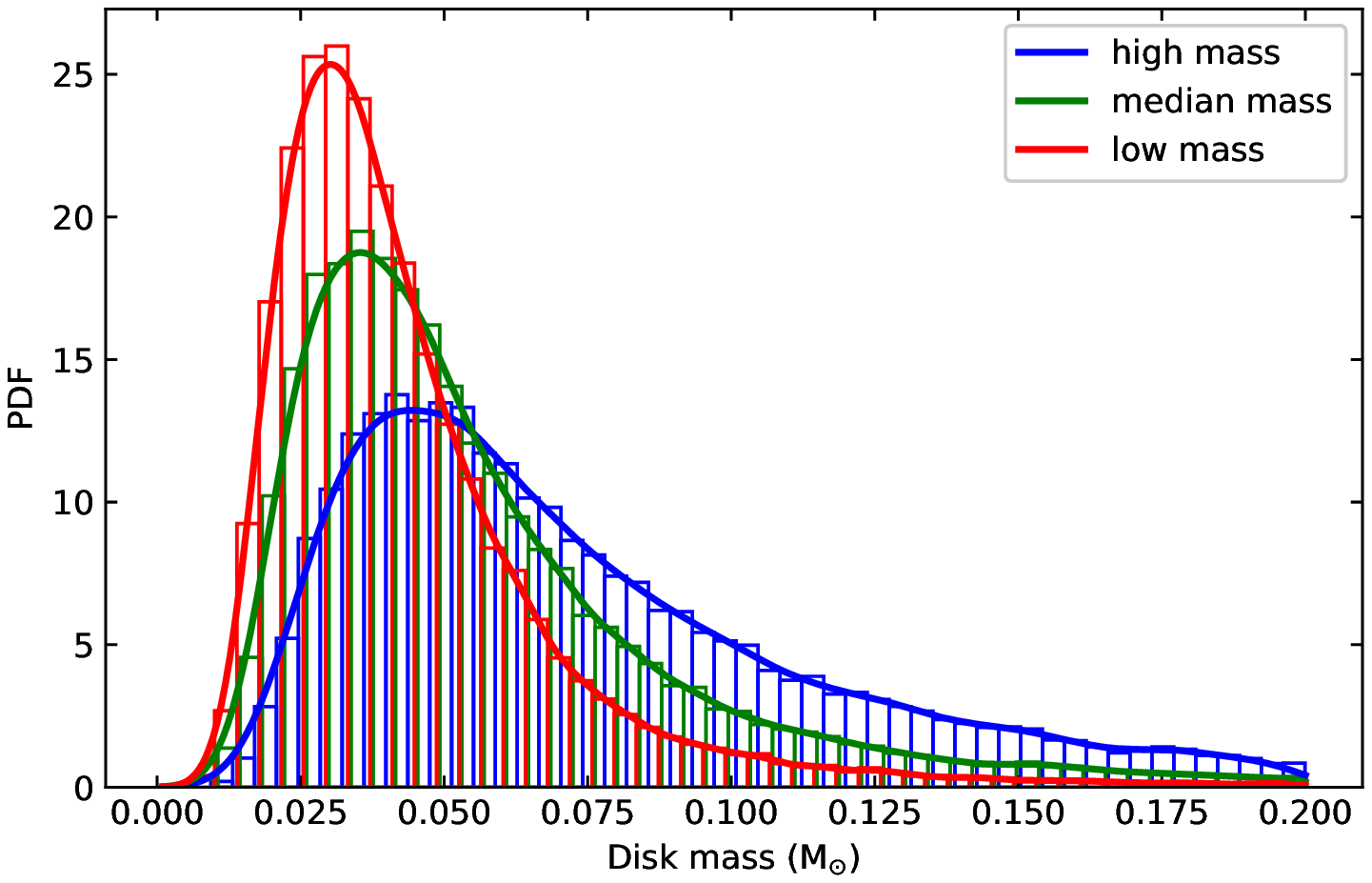}
	\includegraphics[angle=0,scale=0.5]{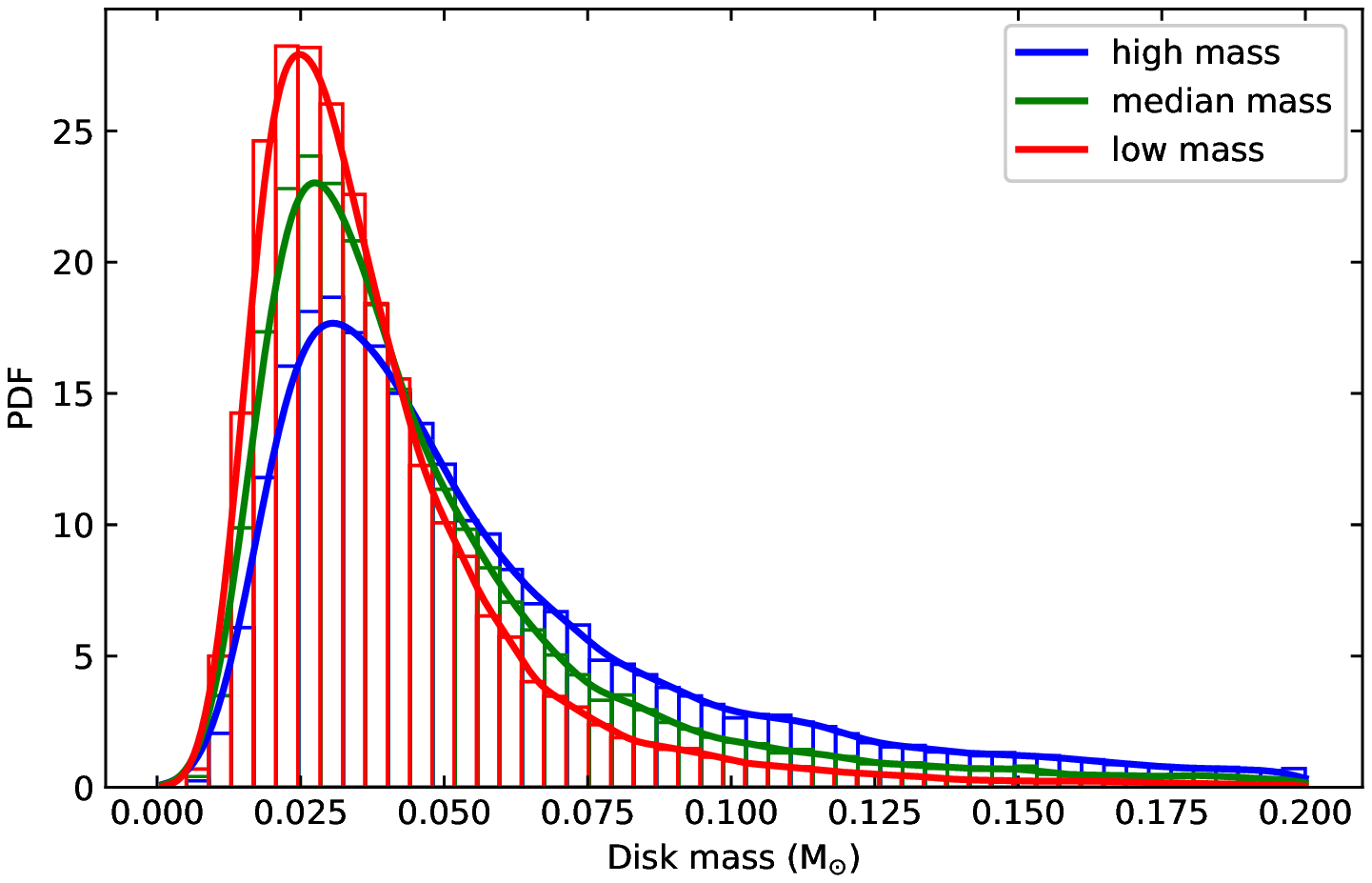}
	\caption{The PDFs of $M_{\rm disk}$ derived from the BZ mechanism. Left: in the case of Gaussian structure; right: in the case of jet+cocoon structure.}
	\hfill
\end{figure}

Our simulation results are summarized in Tab.2 and shown in Fig.\ref{fig:diskmass}. The solid lines are the PDFs for $M_{\rm disk}$ obtained via kernal density estimation on the histograms derived from the simulation, and the three simulation groups are marked with different colors. Combining the results from the groups, the $M_{\rm disk}$ are constrained to be in the range of $0.015-0.134 M_\odot$. Note that the $M_{\rm disk}$ distributions are derived without using the requirement in Eq.(\ref{eq:pmd}), so they are independent of the GW measured $\tilde{\Lambda}$ distribution.

\section{The constraint on EoS}

In this section, we show that the GRB observations are helpful to constrain the $\tilde{\Lambda}$ and hence the EoS models of the neutron stars. Eq.(\ref{eq:md(lambda)}) shows there is a connection between $M_{\rm disk}$ and $\tilde{\Lambda}$. With the $M_{\rm disk}$ distribution inferred from the GRB and kilonova data in Section 2.4, the distribution of $\tilde{\Lambda}$ can be subsequently reconstructed.

\begin{figure}[ht!]
	\figurenum{4}\label{fig:refit}
	\centering
	\includegraphics[angle=0,scale=0.5]{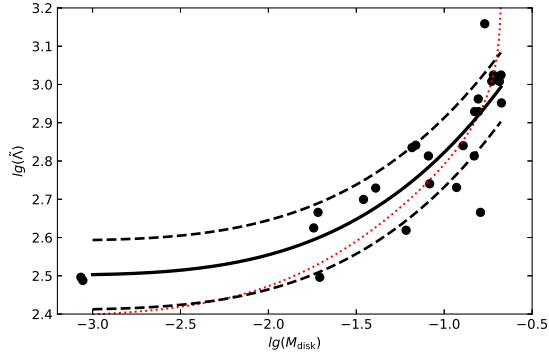}
	\caption{Our best fit of $\tilde{\Lambda}(M_{\rm disk})$ (black solid line) on the simulated data in \citet{2018ApJ...869..130R}, comparing to the inverse function of Eq.(\ref{eq:md(lambda)}) (red dotted line). The standard deviation of the residuals are also shown (black dashed line).}
	\hfill	
\end{figure}

 In Fig.\ref{fig:refit}, we present the simulated data (black circle) in \citet{2018ApJ...869..130R} and the inverse of their empirical relation (red dotted line). Our goal is to express $\tilde{\Lambda}$ as a function of $M_{\rm disk}$, and due to the relatively large dispersion of the data points with respect to Eq.(\ref{eq:md(lambda)}), the best fit on the data that describe $\tilde{\Lambda}(M_{\rm disk})$ might not necessarily be the inverse function of Eq.(\ref{eq:md(lambda)}). For such consideration, we refit the simulated data presented in \citet{2018ApJ...869..130R} (but ignore the points with $M_{\rm disk}$ much lower than $0.001 M_{\odot}$, since they are in the region disfavored by the kilonova modeling) in log space to derive the $\tilde{\Lambda}$ as a function of $M_{\rm disk}$, with the following function
\begin{equation}\label{eq:lam_dm}
\lg(\tilde{\Lambda}) = A_0 + A_1(\lg{M_{\rm disk}}+k)^{A2}.
\end{equation}
By fixing $k = 3.5$, we get $A_0 = 2.502$, $A_1 = 0.013$ and $A_2 = 3.529$, and the standard deviation of the residuals is 0.09. As shown in Fig.\ref{fig:refit}, our best fit (black solid line) is indeed different from the inverse function of Eq.(\ref{eq:md(lambda)}). 

Assuming that $\lg(\tilde{\Lambda})$ has a Gaussian distribution with a mean $\mu(M_{\rm disk})$ given by Eq.(\ref{eq:lam_dm}), the distribution of $lg(\tilde{\Lambda})$ can be obtained by
\begin{equation}
P(\lg\tilde{\Lambda}) =  \int_{M_{\rm d,min}}^{M_{\rm d,max}}P(\tilde{\Lambda}\mid M_{\rm disk})P(M_{\rm disk})dM_{\rm disk},
\end{equation}
where $P(M_{\rm disk})$ can be inferred from the KDE of the simulated $M_{\rm disk}$ samples, and we take $M_{\rm d,min} = 0.001 M_{\odot}$ and $M_{\rm d,max} = 0.2 M_{\odot}$. $P(\tilde{\Lambda}\mid M_{\rm disk})$ can be expressed by
\begin{equation}
P(\tilde{\Lambda}\mid M_{\rm disk}) = \frac{1}{0.09\sqrt{2\pi}}{\rm EXP}\left \{ \frac{[\tilde{\Lambda}-\mu(M_{\rm disk})]^2}{2\cdot 0.09^2} \right \}
\end{equation}
Finally, the distribution of $P(\tilde{\Lambda})$ is given by $P(\tilde{\Lambda}) = P(\lg\tilde{\Lambda})/(\tilde{\Lambda}\ln10)$.

\begin{figure}[ht!]
	\figurenum{5}\label{fig:masslambda}
	\centering
	\includegraphics[angle=0,scale=0.5]{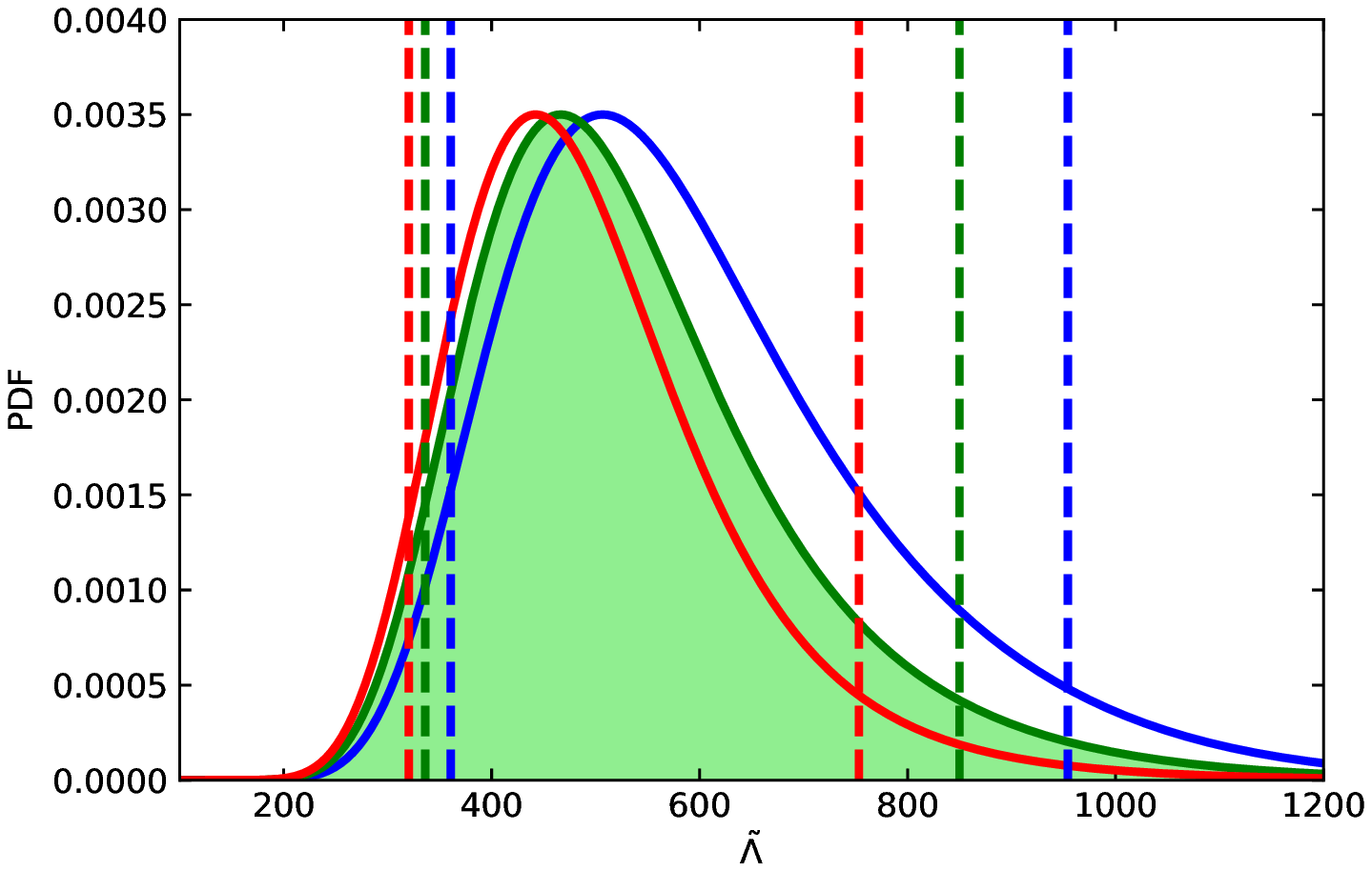}
	\includegraphics[angle=0,scale=0.5]{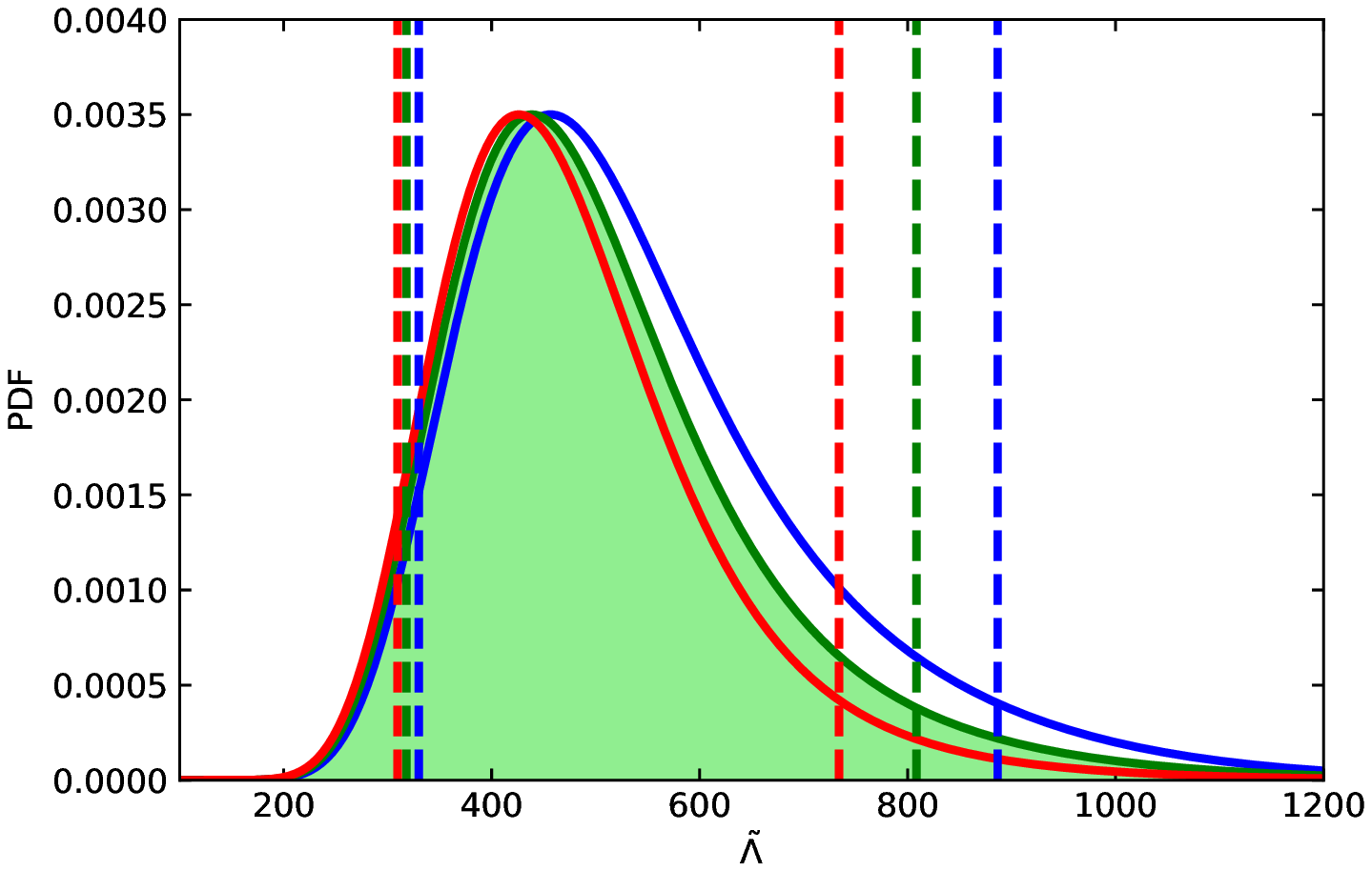}
	\caption{The rescaled PDFs of $\tilde{\Lambda}$ constrained from the outflow energy of GW170817 based on the BZ mechanism. Left: in the case of Gaussian structure; right: in the case of jet+cocoon structure. The vertical dashed lines are the symmetric $90\%$ credible intervals. The results derived from different simulation groups are marked with the same colors as in Fig.\ref{fig:diskmass}.}
	\hfill	
\end{figure}

Following the procedures above, we obtain the distributions of $P(\tilde{\Lambda})$ corresponding to different groups of $M_{\rm disk}$ in Tab.2, and the results are shown in Fig.\ref{fig:masslambda}. An apparent peak apears in the distribution, and for different groups in the simulation, their peaks locate in the range of $\tilde{\Lambda}\sim 420-510$. Taking into account all the structure outflow model and the parameter sets, we have a combined constraint on the tidal deformability $\tilde{\Lambda} \in 309-954$ in $90\%$ confidence level.

The lower bound is set by the kilonova data, in agreement with \citet{2018arXiv181012917R} and \citet{2018arXiv181204803C}, and the upper bound is governed by the GRB data.

\begin{table}
	\label{tab:md_lambda}
	\begin{center}
		\title{}Table 2. The distribution parameters of disk mass and $\tilde{\Lambda}$.\\
		\begin{tabular}{llll} \hline \hline
			Sturcture	&  Group	&  $M_{\rm disk}$  &  $\tilde{\Lambda}$   \\  \hline \hline
		            	&    i      & $0.063^{+0.071}_{-0.037}$ & $568^{+386}_{-208}$ \\
			Gaussian    &    ii     & $0.047^{+0.054}_{-0.026}$ & $513^{+337}_{-177}$ \\
		            	&    iii    & $0.037^{+0.036}_{-0.020}$ & $478^{+276}_{-157}$ \\
			\hline
			            &    i      & $0.046^{+0.069}_{-0.027}$ & $513^{+374}_{-183}$ \\
			Jet+cocoon	&    ii     & $0.037^{+0.051}_{-0.021}$ & $483^{+326}_{-164}$ \\
			            &    iii    & $0.032^{+0.036}_{-0.018}$ & $461^{+273}_{-152}$ \\
			\hline
		\end{tabular}
	\end{center}
	\textbf{Notes.}
	The values mark the median of the distributions, and the errors mark the symmetric $90\%$ credible intervals. \\
\end{table}

Our upper bound is not as tight as that set by the GW observation \citet{2019PhRvX...9a1001A}. This is not surprising since GRB 170817A was viewed off-axis, for which the intrinsic (i.e., if viewed on-axis) $E_\gamma$ and $T_{90}$ are essentially unknown. In the near future, a few on-axis bright sGRBs with GW data are expected to be detected. With the well measured $E_\gamma$, $T_{90}$ and $E_{\rm k}$, our approach will yield significantly tighter constraints.

There are some cautions: first, the reliability of this approach is sensitively dependent of the robustness of Eq.(\ref{eq:md(lambda)}), and as discussed by \citet{2019arXiv190301466K} in the recent paper, the relation needs to be checked for the low mass ratios ($q<0.85$); second, the central remnant is assumed to be a black hole, while the nature of the central remnant of GW170817/GRB 170817A is still in debate.

\section{Discussion}
In this work, assuming a black hole central engine, we have investigated the energy extraction process that launched the relativistic outflow following GW170817. Both the neutrino annihilation mechanism and BZ mechanism have been examined and the uncertainties on the relevant model parameters have been taken into account. Our analysis based on Bayesian inference shows that for $T_{\rm act} \sim 2 \rm{s}$, the BZ mechanism is more favored, since the neutrino annihilation model is more likely to predict a disk mass which exceeds the limitation from gravitational wave observation. On the other hand, if $T_{\rm act}$ is much (an order of magnitude) smaller than the duration of the observed gamma-ray emission, these two models become indistinguishable.

We also derived the disk mass distribution for BZ mechanism and the mass of the torus around the central black hole is estimated to be within the range of $0.015-0.134 M_\odot$. If the disk mass is large enough within probability distribution estimated from BZ mechanism, for example $M_{\rm disk}\sim 0.1 M_\odot$ and the disk wind carries away $\sim 20\%$, then $\sim 0.02 M_\odot$ of material in the wind may be capable to produce the blue component of the kilonova, making an unified physical picture to explain the EM counterparts of GW170817.
We further derived the probability distribution for $\tilde{\Lambda}$. The $90\%$ credible interval of $\tilde{\Lambda}$ for the progenitor of GW170817 is found to be $309-954$. One thing should be pointed out is that GRB 170817A is an under-luminous/off-axis event, for which the lack of a reliable measurement of both the intrinsic $E_\gamma$ and $T_{90}$ adds sizeable uncertainties to our result. This shortcoming will be overcome for the future GW events associated with bright on-axis GRBs. The rate of such association events is expected to be $\sim 1\pm0.5$ event/year for a full-sky like gamma-ray monitor in the full-sensitivity run era of advanced LIGO/Virgo \citep{LiX2016ApJ} and a sample will be established in the next decade. For such events, if clear breaks in the afterglow light curves have been well recorded, then the total energy of GRB's outflow and afterglow can be reliably inferred. With our approach outlined in this work, the constraints on the energy extraction process as well as the EoS of neutron stars (i.e., $\tilde{\Lambda}$) will be significantly enhanced, benefited from also the expected improvements on the correlation between $M_{\rm disk}$ and $\tilde{\Lambda}$ and on understanding the nature of central remnant. Further efforts could be made to improve our method. To make full use of the information in multi-messanger observations, joint parameter estimation of GW and GRB as well as the kilonova should be made (although for now this would be very computational expensive). The accuracies of astrophysical models should be improve to make reliable constraints on the parameters. when this work is under revision, \citet{2018arXiv181204803C} proposed an new relation among $M_{\rm disk}$, $M_{\rm tot}$, and $M_{\rm thr}$, and used it to constrain $\tilde{\Lambda}$ and $q$. More studies should be made to put forward and check the robustness of these kind of relations. at last, the distributions of parameters obtained from the modeling of EM counterparts are crucial to this kind of works, and we appeal to the community to share the posterior samples if relevant parameters are inferred.

\clearpage
\section*{Acknowledgments}
We thank the anonymous referee for the suggestions that are helpful to improve our work. This work was supported in part by NSFC under grants of No. 11525313 (i.e., Funds for Distinguished Young Scholars), No. 11433009, No. 11773078 and 11763003, the Foundation for Distinguished Young Scholars of Jiangsu Province (No. BK20180050), the Chinese Academy of Sciences via the Strategic Priority Research Program (Grant No. XDB09000000) and and the International Partnership Program of Chinese Academy of Sciences (114332KYSB20170008). F.-W.Z. also acknowledges the support by the Guangxi Natural Science Foundation (No. 2017GXNSFAA198094).

\clearpage

\end{document}